\documentclass[letter,aps,superscriptaddress,nofootinbib,tightenlines,onecolumn,prl]{revtex4}

\usepackage{amsfonts}
\usepackage{multirow}
\usepackage{mathrsfs}
\usepackage{graphicx}
\usepackage{amsmath}
\usepackage{amssymb}
\usepackage{bm}
\usepackage{bbm}
\usepackage{color}
\usepackage{array}
\usepackage{diagbox}
\usepackage{float}
\usepackage{makecell}

\def\OMIT#1{}

\def\hlinew#1{%
  \noalign{\ifnum0=`}\fi\hrule \@height #1 \futurelet
   \reserved@a\@xhline}
\usepackage{array}
\newcommand{\PreserveBackslash}[1]{\let\temp=\\#1\let\\=\temp}
\newcolumntype{C}[1]{>{\PreserveBackslash\centering}p{#1}}
\newcolumntype{R}[1]{>{\PreserveBackslash\raggedleft}p{#1}}
\newcolumntype{L}[1]{>{\PreserveBackslash\raggedright}p{#1}}

\newcommand{\nn}{\nonumber}

\newcommand{\beq}{\begin{equation}}
\newcommand{\eeq}{\end{equation}}
\newcommand{\bqa}{\begin{eqnarray}}
\newcommand{\eqa}{\end{eqnarray}}

\newcommand\fverb{\setbox\fverbbox=\hbox\bgroup\verb}
\newcommand\fverbdo{\egroup\medskip\noindent%
			\fbox{\unhbox\fverbbox}\ }
\newcommand\fverbit{\egroup\item[\fbox{\unhbox\fverbbox}]}
\newbox\fverbbox

\makeatletter

\newcommand{\Rmnum}[1]{\expandafter\@slowromancap\romannumeral #1@}
\makeatother

\begin{document}
\title{\mbox{}\\[10pt]
Higher-order QCD corrections to $\Upsilon$ decay into double charmonia}
\author{Yu-Dong Zhang~\footnote{zhangyudong@email.swu.edu.cn}}
\affiliation{School of Physical Science and Technology, Southwest University, Chongqing 400700, China}

\author{Wen-Long Sang~\footnote{wlsang@swu.edu.cn}}
\affiliation{School of Physical Science and Technology, Southwest University, Chongqing 400700, China}

\author{Hong-Fei Zhang~\footnote{shckm2686@163.com}}
\affiliation{College of Big Data Statistics, Guizhou University of Finance and Economics, Guiyang, 550025, China}

\date{\today}
\begin{abstract}
In this work, we study the exclusive decay of $\Upsilon$ into $J/\psi$ in association with $\eta_c$ ($\chi_{c0,1,2}$).
The decay widths for different helicity configurations are evaluated up to QCD next-to-leading order within the nonrelativistic QCD framework.
We find that the QCD corrections notably mitigate the renormalization scale dependence of the decay widths for all the processes.
The branching fraction of $\Upsilon\rightarrow J/\psi+\chi_{c1}$ is obtained as $3.73^{+5.10+0.10}_{-2.06-1.19}\times 10^{-6}$,
which agrees well with the Belle measurement, i.e., ${\rm Br}(\Upsilon\rightarrow J/\psi+\chi_{c1})=(3.90\pm1.21\pm0.23)\times10^{-6}$.
For the other processes, our results of the branching fractions are compatible with the upper limits given by the Belle experiments,
except for $\Upsilon(2S)\to J/\psi +\chi_{c1}$, where some tension exists between theory and experiment.
Having the polarized decay widths, we study the $J/\psi$ polarization, which turn out to be independent of any nonperturbative parameters.
Further, according to our calculation, it is promising to measure all the processes at Super B-factory thanks to the high luminosity.

\end{abstract}
\maketitle


The exclusive decay of a bottomonium into double charmonia provides an excellent probe into the nonrelativistic QCD (NRQCD) factorization~\cite{Bodwin:1994jh},
since all the involved external particles are heavy quarkonia and thus the uncertainties originating from any other nonperturbative parameters are screened out,
except for only those from the long-distance matrix elements.
To date, many of such processes have been intensively studied, including $\eta_b\to J/\psi +J/\psi$~\cite{Jia:2006rx,Gong:2008ue,Braguta:2009xu,Sun:2010qx},
$\chi_{bJ}\to J/\psi +J/\psi$~\cite{Braguta:2005gw,Braguta:2009df,Zhang:2011ng,Sang:2011fw,Chen:2012ih},
$\Upsilon \to J/\psi+\eta_c$~\cite{Jia:2007hy,Sang:2015owa}, and $\Upsilon \to J/\psi+\chi_{cJ}$~\cite{Xu:2012uh}.
On the experimental side, the {\tt Belle} collaboration has collected enormous $\Upsilon(1S)$ and $\Upsilon(2S)$ events,
and measured the branching fraction ${\rm Br}(\Upsilon(1S)\to J/\psi+\chi_{c1})=3.90\pm1.21(\rm stat.)\pm0.23(\rm syst.) \times 10^{-6}$~\cite{Belle:2014wam}.
For some other processes, such as $\Upsilon\to J/\psi+\chi_{c0,2}$ and $\Upsilon\to J/\psi+\eta_c$,
the upper limits of the branching fractions are determined.
The QCD leading order (LO) results for these processes given in Refs.~\cite{Jia:2007hy, Sang:2015owa, Xu:2012uh} suffer from significant uncertainties.
For instance, as the renormalization scale $\mu_R$ runs from twice of the charm quark mass, $2m_c$, to twice of the bottom quark mass, $2m_b$,
the decay width changes by a typical factor of 5,
which destroys the predicting power of the phenomenological results.
In order to confront the theory with data, it urgent to tackle the computation of higher-order QCD corrections.

In recent years, technological advances have made it possible to calculate the higher-order QCD corrections to the processes involving two external quarkonia.
In Ref.~\cite{Zhang:2021ted}, the decay width of $\Upsilon\to \eta_c(\chi_{cJ})+\gamma$ was evaluated up to QCD next-to-leading order (NLO),
and the very challenging two-loop perturbative corrections to $e^+e^-\to J/\psi+\eta_c(\chi_{cJ})$ was given in Refs.~\cite{Feng:2019zmt,Sang:2022kub}.
For all these processes, the renormalization scale dependences become much milder at two-loop level than those at LO,
which indicates good convergence of the perturbative expansion.
With all the available perturbative corrections lumped together,  
the theoretical results on the cross sections of $J/\psi+\eta_c(\chi_{c0})$ production agree with the experimental measurements, notwithstanding large uncertainties. 

\begin{figure}[htbp]
\centering
\includegraphics[width=0.5\textwidth]{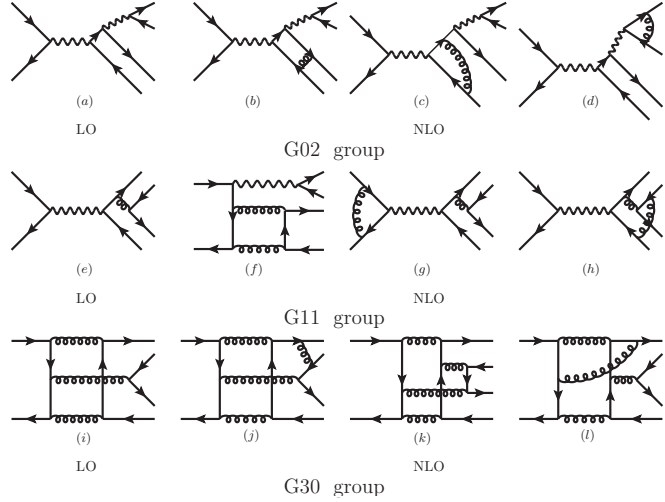}
\caption{
Some representative Feynman diagrams, drawn with \texttt{JaxoDraw}~\cite{Binosi:2008ig},
for $\Upsilon\to J/\psi+\eta_c (\chi_{cJ})$.
\label{fig-feynman-diagram}}
\end{figure}

Inspired by the success in the processes aforementioned, we calculate, in the current work, the NLO QCD corrections to $\Upsilon\to J/\psi+\eta_c(\chi_{cJ})$, which involves three external quarkonia.
At lowest order in $\alpha_s$, the Feynman diagrams of these processes can be classified into three groups,
as illustrated in Fig.~\ref{fig-feynman-diagram}(a),~\ref{fig-feynman-diagram}(e), and \ref{fig-feynman-diagram}(i),
the amplitudes of which are proportional to $\alpha^2$, $\alpha_s\alpha$, and $\alpha_s^3$, respectively. Here $\alpha_s$ and 
$\alpha$ are the strong coupling and electromagnetic coupling, respectively.  
In the following, we denote these three groups by the symbols ${\rm G02}$, ${\rm G11}$, and ${\rm G30}$, respectively,
where the two numbers represent the powers of their amplitudes in $\alpha_s$ and $\alpha$ respectively. 
In the ${\rm G02}$ group, the $J/\psi$ meson is produced via a photon fragmentation,
which may enhance the amplitudes by a factor of $m_b^2/m_c^2$ relative to the non-fragmentation diagrams.
As one can see in Fig.~\ref{fig-feynman-diagram}, there does not exist tree-level diagrams in the ${\rm G30}$ group,
owing to the charge parity and color conservation,
accordingly, its NLO corrections are two-loop diagrams with six external quark lines, the evaluation of which is extremely difficult.

The differential decay width of a $\Upsilon$ with polarization (along the $z$-axis) $S_z$ into a $J/\psi$ and another charmonium, $H$ (here, $H$ is either an $\eta_c$ or a $\chi_{cJ}$ meson),
the helicities of which are $\lambda_1$ and $\lambda_2$, respectively, can be expressed as~\cite{Haber:1994pe,Jacob:1959at}
\bqa\label{eq-gen-rate-helicity}
\frac{\mathrm{d}\Gamma}{\mathrm{d}\cos\theta}(\Upsilon(S_{z}) \rightarrow J / \psi(\lambda_1)+H(\lambda_2)) 
=\frac{|\mathbf{P}|}{16 \pi m_{\Upsilon}^{2}}\left|d_{S_{z}, \lambda_1-\lambda_2}^{1}(\theta)\right|^{2}\left|\mathcal{A}_{\lambda_1, \lambda_2}^{H}\right|^{2},
\eqa
where $m_\Upsilon$ is the mass of the $\Upsilon$ meson, $\mathbf{P}$ denotes the spatial components of the $J/\psi$ momentum,
$\mathcal{A}_{\lambda_1, \lambda_2}^{H}$ is the Feynman amplitude corresponding to the helicity configuration ($\lambda_1$, $\lambda_2$),
and $d^1_{S_z,\lambda_1-\lambda_2}(\theta)$ is the Wigner function.
Here, $\theta$ is the angle between the direction of $\mathbf{P}$ and the $z$-axis.
Note that the constraint, $\lambda_{1}-\lambda_{2}\leq 1$, is guaranteed by the angular momentum conservation.

Integrating over the polar angle $\theta$ and averaging over the polarization of $\Upsilon$,
we finally obtain the integrated decay width of $\Upsilon\to J/\psi+H$  for the helicity configuration ($\lambda_{1}$,$\lambda_{2}$) as
\begin{equation}\label{eq-gen-rate-helicity-int}
\Gamma(\Upsilon \rightarrow J/\psi(\lambda_1)+H(\lambda_{2}))=\frac{|\mathbf{P}|}{24\pi M_{\Upsilon}^{2}}\left|\mathcal{A}_{\lambda_1,\lambda_2}^{H}\right|^{2}.
\end{equation}
Thanks to the parity invariance, we have the following relations,
\bqa\label{eq-helicity-parity-invariance}
\mathcal{A}^{\eta_{c}}_{\lambda_1,\lambda_2}=-\mathcal{A}^{\eta_{c}}_{-\lambda_1,-\lambda_2},\quad
\mathcal{A}^{\chi_{cJ}}_{\lambda_1,\lambda_2}=(-1)^J \mathcal{A}^{\chi_{cJ}}_{-\lambda_1,-\lambda_2},
\eqa
and the number of independent helicity amplitudes for $\eta_c$, $\chi_{c0}$, $\chi_{c1}$,
and $\chi_{c2}$ production can be further reduced to one, two, three, and five, respectively.
In terms of the independent helicity amplitudes, the unpolarized decay widths can be explicitly written as
\begin{subequations}\label{eq-gen-rate-helicity-explicit}
\begin{eqnarray}
&&\Gamma(\Upsilon\to J/\psi+\eta_{c})=
\frac{|{\bf P}|}{24\pi m_\Upsilon^{2}}\bigg(2| A_{1,0}^{\eta_{c}}|^2\bigg),\\
&&\Gamma(\Upsilon\to J/\psi+\chi_{c0}) =\frac{|{\bf P}|}{24\pi m_\Upsilon^{2}}\bigg(2|A_{1,0}^{\chi_{c0}}|^2+|A_{0,0}^{\chi_{c0}}|^2\bigg),\\
&&\Gamma(\Upsilon\to J/\psi+\chi_{c1}) =\frac{|{\bf P}|}{24\pi m_\Upsilon^{2}}\bigg(2| A_{1,1}^{\chi_{c1}}|^2+2| A_{1,0}^{\chi_{c1}}|^2+2| A_{0,1}^{\chi_{c1}}|^2\bigg),\\
&&\Gamma(\Upsilon\to J/\psi+\chi_{c2})=\frac{|{\bf P}|}{24\pi m_\Upsilon^{2}}\bigg(2| A_{1,2}^{\chi_{c2}}|^2+2|A_{1,1}^{\chi_{c2}}|^2+2|A_{1,0}^{\chi_{c2}}|^2+2|A_{0,1}^{\chi_{c2}}|^2+|A_{0,0}^{\chi_{c2}}|^2\bigg).
\end{eqnarray}
\end{subequations}

According to the NRQCD factorization formalism~\cite{Bodwin:1994jh,Zhang:2021ted}, 
${\mathcal A}_{\lambda_1,\lambda_2}^{H}$ can be factorized as
\bqa\label{eq-nrqcd}
&&{\mathcal A}_{\lambda_1,\lambda_2}^{H}=\sqrt{2m_{\Upsilon}}\sqrt{2m_H}\sqrt{2m_{J/\psi}}\mathcal{C}_{\lambda_1,\lambda_2}^H\frac{N_c R_{\Upsilon}(0)R_{J/\psi}(0)}{2\pi m_c^{3/2}m_b^{3/2}} \nn \\
&&~~\times\begin{cases}\sqrt{\frac{N_c}{2\pi}}\frac{R_{\eta_c}(0)}{m_c^{3/2}}, &\text{for}\;\eta_c \\
\sqrt{\frac{3 N_c}{2\pi}}\frac{R^\prime_{\chi_{cJ}}(0)}{ m_c^{5/2}}, & \text{for}\;\chi_{cJ}\end{cases}
\eqa
where $N_c=3$ is the number of colors, $m_{J/\psi}$ and $m_H$ are the mass of $J/\psi$ and $H$, respectively,
and $R(0)$ and $R^\prime(0)$ denote the $S$-wave radial wave functions at the origin and the first derivative of the $P$-wave radial wave function at the origin, respectively.
Exploiting the heavy quark spin symmetry, we can make the following approximations, ${R_{J/\psi}}(0)\approx{R_{\eta_c}}(0)$ and ${R^{\prime}_{\chi_{c0}}}(0)\approx{R^{\prime}_{\chi_{c1}}}(0)\approx{R^{\prime}_{\chi_{c2}}}(0)$.
In the above equation, $\mathcal{C}_{\lambda_1,\lambda_2}^H$ denotes the short-distance coefficient (SDC) of the corresponding helicity amplitude,
and satisfies the helicity selection rule~\cite{Chernyak:1980dj,Brodsky:1981kj},
\bqa\label{eq-helicity-scaling-rule-1}
{\mathcal C}_{\lambda_1,\lambda_2}^H\propto \bigg(\frac{m_c}{m_b}\bigg)^{1+|\lambda_{1}+\lambda_{2}|},
\eqa
in the asymptotic limit $m_c/m_b\ll 1$.
All these SDCs can be determined following the standard perturbative matching precedures.

It is convenient to decompose each SDC as
\begin{eqnarray}
\label{eq-sdcs}
\mathcal C_{\lambda_{1}, \lambda_{2}}^H=\mathcal C^{H,{\rm G02}}_{\lambda_{1}, \lambda_{2}}+\mathcal C^{H,{\rm G11}}_{\lambda_{1}, \lambda_{2}}+\mathcal C^{H,{\rm G30}}_{\lambda_{1}, \lambda_{2}},
\end{eqnarray}
where the three SDCs on the right-hand side correspond to the contributions from $\rm G02$, $\rm G11$, and $\rm G30$ groups, respectively. 
Up to NLO, these SDCs can be expanded in powers of the coupling constants and formally written as
\bqa\label{eq-sdcs-expand}
&&\mathcal C^{H,{\rm G}}_{\lambda_{1}, \lambda_{2}}=\alpha_s^m\alpha^n\bigg[\mathcal C^{H,{\rm G},(0)}_{\lambda_{1}, \lambda_{2}}+\frac{\alpha_s}{\pi}\times\bigg(\frac{m}{4}\beta_0 \ln\frac{\mu_R^2}{m_b^2}\, \mathcal C^{H,{\rm G},(0)}_{\lambda_{1}, \lambda_{2}}+\mathcal C^{H,{\rm G},(1)}_{\lambda_{1}, \lambda_{2}}\bigg)+\mathcal{O}(\alpha_s^2)\bigg],
\eqa
where ($m$, $n$)=(0, 2), (1, 1), and (3, 0) for $\rm G02$, $\rm G11$, and $\rm G30$, respectively.
Here, $\beta_0=(11/3)C_A-(4/3)T_Fn_f$ is the one-loop coefficient of the QCD $\beta$ function,
where $T_F=\tfrac{1}{2}$ and $n_f$ is the number of active quark flavors.
In this work, we take $n_f=n_L+n_H$, where $n_L=3$ and $n_H=1$ are the number of light quark and heavy quark flavors, respectively.

The quark-level Feynman diagrams and Feynman amplitudes are generated using {\tt FeynArts}~\cite{Hahn:2000kx}.
Employing the color and spin projectors followed by enforcing spin-orbit coupling,
we obtain the hadron-level amplitudes order by order in $\alpha_s$ with the aid of the packages {\tt FeynCalc}~\cite{Shtabovenko:2016sxi} and {\tt FormLink}~\cite{Feng:2012tk}.
Note that the helicity projectors constructed in Refs.~\cite{Jia:2007hy,Xu:2012uh} are adopted to evaluate the polarized amplitudes.
Utilizing the packages {\tt Apart}~\cite{Feng:2012iq} and {\tt FIRE}~\cite{Smirnov:2014hma}, we can further reduce the loop integrals into linear combinations of master integrals (MIs). 
Finally, we end up with 36 one-loop MIs, which, in our calculation, are evaluated using {\tt Package-X}~\cite{Patel:2015tea},
and roughly 1800 two-loop MIs, the evalutaion of which is a challenging work.
Fortunately, a powerful new algorithm, dubbed Auxiliary Mass Flow (AMF), has recently been pioneered by Liu and Ma~\cite{Liu:2017jxz,Liu:2020kpc,Liu:2021wks}.
Its main idea is to set up differential equations with respect to an auxiliary mass variable, with the vacuum bubble diagrams as the boundary conditions.
Remarkably, these differential equations can be solved iteratively with very high numerical precision.
To tackle the complicated MIs with high precision in the previous work investigating the process, $e^+e^-\to J/\psi+\chi_{cJ}$,
we have applied the AMF approach which turns out to be highly efficient in the numerical evaluation of the multi-loop integrals.
Therefore, in this work, we utilize the newly released package {\tt AMFlow}~\cite{Liu:2022chg} to compute all the two-loop MIs. 
After implementing renormalization of the heavy quark mass and field strength, and the QCD coupling,
where the on-shell and $\overline{\mathrm{MS}}$ renormalization schemes are employed, respectively,
we numerically verify that all the UV poles indeed cancel with high precision.

In our numerical computation, we take the $S$-wave radial wave function at the origin and first derivative of the $P$-wave radial wave function at the origin from Ref.~\cite{Eichten:1995ch} as
\begin{subequations}
\begin{eqnarray}
&\left|R_{\Upsilon(1S)}(0)\right|^{2}=6.477 \mathrm{GeV}^{3},  \\
&\left|R_{\Upsilon(2S)}(0)\right|^{2}=3.234 \mathrm{GeV}^{3},  \\
&\left|R_{\Upsilon(3S)}(0)\right|^{2}=2.474 \mathrm{GeV}^{3}, \\
&\left|R_{J/\Psi/\eta_c}(0)\right|^{2}=0.81 \mathrm{GeV}^{3},  \\
&\left|R_{\eta_c(2S)}(0)\right|^{2}=0.529 \mathrm{GeV}^{3},  \\
&\left|R_{\chi_{cJ}}^{\prime}(0)\right|^{2}=0.075 \mathrm{GeV}^{5},
\end{eqnarray}
\end{subequations}
which were obtained based on the Buchm\"uller-Tye potential model.
Since the relativistic corrections are not considered, we are allowed to make the following approximations,
$m_\Upsilon\approx 2 m_b$, $m_{J/\psi}\approx 2m_c$, and $m_H\approx 2 m_c$.
The value of the QED running coupling at the mass of the $\Upsilon$ meson is set to be $\alpha(m_\Upsilon)=\tfrac{1}{131}$,
while the values of the QCD running coupling at various renormalization scales are computed with the aid of the package {\tt RunDec}~\cite{Chetyrkin:2000yt}.
In order to calculate the branching fractions, we take the values of the total decay widths of $\Upsilon(\mathrm{nS})$ from the latest particle data group (PDG)~\cite{ParticleDataGroup:2020ssz},
explicitly, $\Gamma_{\Upsilon(1S)}=54.02\pm 1.25\, {\rm keV}$, $\Gamma_{\Upsilon(2S)}=31.98\pm 2.63\, {\rm keV}$ and $\Gamma_{\Upsilon(3S)}=20.32\pm 1.85\, {\rm keV}$.

Since the theoretical results of the decay width depend on the values of the heavy quark mass and renormalization scale $\mu_R$,
we vary $m_c$ from $1.3~\mathrm{GeV}$ to $1.7~\mathrm{GeV}$, $m_b$ from $4.4~\mathrm{GeV}$ to $4.8~\mathrm{GeV}$,
and $\mu_R$ from $2m_c$ to $2m_b$, in order to estimate the theoretical uncertainties.
Note here that the range of $\mu_R$ is fixed for each specific choice of the $m_c$ and $m_b$ values.
In addition, we will present results at $m_c=1.5~\mathrm{GeV}$, $m_b=4.6~\mathrm{GeV}$, and $\mu_R=m_b$ as the so-called central values.

Having completed our calculation, we compare the results with those in the existing literature as a partial check.
By taking the same parameters,
our results on the LO SDCs of $\Upsilon\to J/\psi+\eta_c$ and $\Upsilon\to J/\psi+\chi_{cJ}$ are consistent with those in Ref.~\cite{Jia:2007hy} and Ref.~\cite{Xu:2012uh}, respectively.
In addition to the LO diagrams, these two references also presented results for the diagrams corresponding to type 
(f) in Fig.~\ref{fig-feynman-diagram}.
Our results for this group of diagrams are consistent with those in Ref.~\cite{Jia:2007hy}, while do not agree with those in Ref.~\cite{Xu:2012uh}.
As an interesting observation, in our results the imaginary part of the amplitude for $\Upsilon\to J/\psi+\chi_{c1}$ vanishes,
while that in Ref.~\cite{Xu:2012uh} does not.
As a matter of fact, this phenomenon can be accessed through a qualitative analysis.
Invoking the Cutkosky rule, one can find this imaginary part is proportional to the amplitude of the process, $\chi_{c1}\to2{\rm gluons}$,
which is strictly forbidden according to the Landau-Yang theorem.

Having obtained the numerical results for all the amplitudes,
We find that the $\mathrm{G30}$, $\mathrm{G11}$, and $\mathrm{G02}$ groups makes the most, second, and last important contributions,
which confirms the necessity of doing the challenging two-loop computations.

\begin{table*}[!htbp]
\caption{
Theoretical results of the decay widths ($10^{-3}\rm eV$) for $\Upsilon\to J/\psi+\eta_c(\chi_{cJ})$ at LO and NLO.
}
\label{tab-decay-rate}
\centering
\renewcommand{\arraystretch}{1.2}
\begin{tabular}{|c|c|c|c|c|c|c|c|c|}
\hline
$\rm H$&$\rm Order$&$(0,0)$&$(1,0)$&$(0,1)$&$(1,1)$&$(1,2)$&$\rm \Gamma_{Unpol}$
\\
\hline
\multirow{2}*{$\eta_{c}$}
&$\rm LO$&$-$&$86.90^{+85.63}_{-42.71}$&$-$&$-$&$-$&$173.81^{+171.26}_{-85.43}$
\\
\cline{2-8}
&$\rm NLO$&$-$&$35.97^{+51.89}_{-21.18}$&$-$&$-$&$-$&$71.93^{+103.78}_{-42.37}$
\\
\hline
		\multirow{2}*{$\eta_{c}(2S)$}
		&$\rm LO$&$-$&$56.72^{+55.91}_{-27.82}$&$-$&$-$&$-$&$113.44^{+111.82}_{-55.64}$
		\\
		\cline{2-8}
		&$\rm NLO$&$-$&$23.49^{+33.76}_{-13.78}$&$-$&$-$&$-$&$47.00^{+67.53}_{-27.55}$
		\\
		\hline
\multirow{2}*{$\chi_{c0}$}
&$\rm LO$&$13.97^{+8.09}_{-5.27}$&$31.27^{+32.66}_{-15.26}$&$-$&$-$&$-$&$76.50^{+73.40}_{-35.79}$
\\
\cline{2-8}
&$\rm NLO$&$12.84^{+10.57}_{-5.91}$&$30.37^{+43.55}_{-17.20}$&$-$&$-$&$-$&$73.59^{+97.67}_{-40.31}$
\\
\hline
\multirow{2}*{$\chi_{c1}$}
&$\rm LO$&$-$&$62.44^{+84.56}_{-34.18}$&$28.33^{+28.30}_{-13.54}$&$19.38^{+15.59}_{-8.40}$
&$-$&$220.31^{+256.89}_{-112.24}$
\\
\cline{2-8}
&$\rm NLO$&$-$&$49.60^{+79.93}_{-29.33}$&$31.71^{+38.45}_{-16.83}$&$19.34^{+19.49}_{-9.51}$
&$-$&$201.30^{+275.73}_{-111.33}$
\\
\hline
\multirow{2}*{$\chi_{c2}$}
&$\rm LO$&$2.04^{+3.69}_{-1.15}$&$6.56^{+10.80}_{-3.95}$&$1.84^{+1.61}_{-0.83}$
&$2.61^{+2.53}_{-1.25}$&$1.49^{+0.77}_{-0.58}$&$27.03^{+35.11}_{-14.37}$
\\
\cline{2-8}
&$\rm NLO$&$0.66^{+1.57}_{-0.33}$&$6.73^{+15.22}_{-4.59}$&$1.42^{+1.69}_{-0.72}$
&$2.97^{+4.20}_{-1.72}$&$1.59^{+1.39}_{-0.74}$&$26.09^{+46.56}_{-15.87}$
\\
\hline
\end{tabular}
\end{table*}

In Tab.~\ref{tab-decay-rate}, we present our results for not only the unpolarized decay widths,
but also the polarized decay widths for each independent helicity configuration,
where the uncertainties are obtained by varying the values of $m_c$ and $m_b$.
 In the current work, we do not take into account the relativistic corrections, which may potentially bring about an extra uncertainty.
We can naively estimate it as about $40\%$ of the leading order results according to the scaling, $v^2\sim0.3$ for charmonium and $v^2\sim0.1$ for bottomonium,
where $v$ is the typical velocity of the heavy quarks in a quarkonium.
For almost all the processes, the NLO corrections are negative and moderate,
which usually indicates good convergence of the perturbative expansion.
As exceptions, the NLO corrections to the $(0,1)$ channel of $J/\psi+\chi_{c1}$ is positive,
and those to $J/\psi+\eta_c$ and the $(0,0)$ channel of $J/\psi+\chi_{c2}$ have a pronounced impact. 

Having the results with helicity configurations, the $J/\psi$ polarization turns out to be an interesting physical observable.
As a well-known puzzle, the $J/\psi$ polarization at hadron colliders still cannot be well described by the NRQCD computation at NLO.
Our current work provides an ideal laboratory for the study of the $J/\psi$ polarization,
since all the nonperturbative parameters cancel in the polarization parameters.
Among these parameters, $\lambda_\theta$, defined as
\bqa
\lambda_\theta=\frac{\Gamma_T-\Gamma_L}{\Gamma_T+\Gamma_L},
\eqa
is the one that attracts the widest attention.
Here, $\Gamma_T$ and $\Gamma_L$ are the decay widths in Eq.~(\ref{eq-gen-rate-helicity-int}) for $\lambda_1=1$ and 0, respectively.
In the case $\mathrm{H}=\eta_c$, the $J/\psi$ is transversely polarized, thus we have $\lambda_\theta(\Upsilon\to J/\psi+\eta_c)=1$.
For $\mathrm{H}=\chi_c$, the values of $\lambda_\theta$ at QCD NLO are
\begin{subequations}
\bqa
&&\lambda_\theta(\Upsilon\to J/\psi+\chi_{c0})=0.41^{+0.12+0.04}_{-0.10-0.00},  \\
&&\lambda_\theta(\Upsilon\to J/\psi+\chi_{c1})=0.04^{+0.05+0.08}_{-0.04-0.15},  \\
&&\lambda_\theta(\Upsilon\to J/\psi+\chi_{c2})=0.52^{+0.06+0.01}_{-0.11-0.30},  \\
&&\lambda_\theta(\Upsilon\to J/\psi+\chi_c)=0.16^{+0.07+0.05}_{-0.06-0.13}, \label{eqn:pol}
\eqa
\end{subequations}
where the two uncertainties are from the choices of the heavy quark mass and the renomalization scale.
In the evaluation of the last quantity in Eq.~(\ref{eqn:pol}), $\Gamma_T$ ($\Gamma_L$) sums over all the contributions from $\chi_{c0}$, $\chi_{c1}$, and $\chi_{c2}$.
It is interesting to note that the uncertainties from the ambiguities of the heavy quark mass are considerably reduced. 

\begin{figure}[htbp]
\centering
\includegraphics[width=0.4\textwidth]{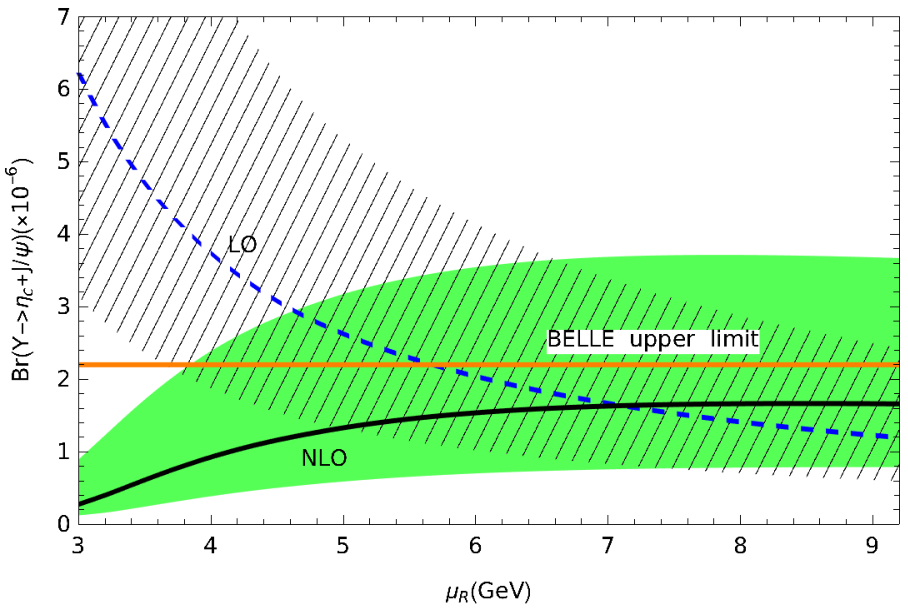}
\includegraphics[width=0.4\textwidth]{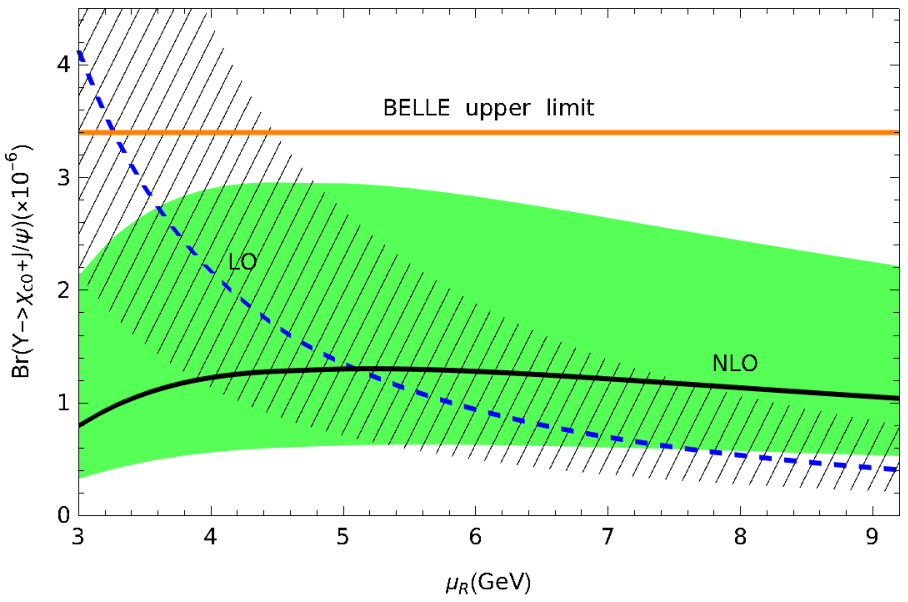}
\includegraphics[width=0.4\textwidth]{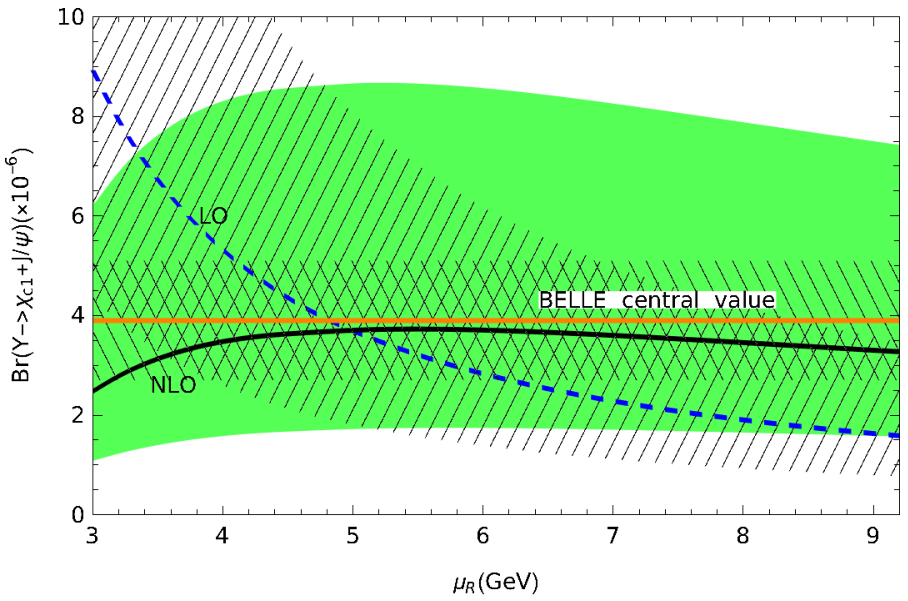}
\includegraphics[width=0.4\textwidth]{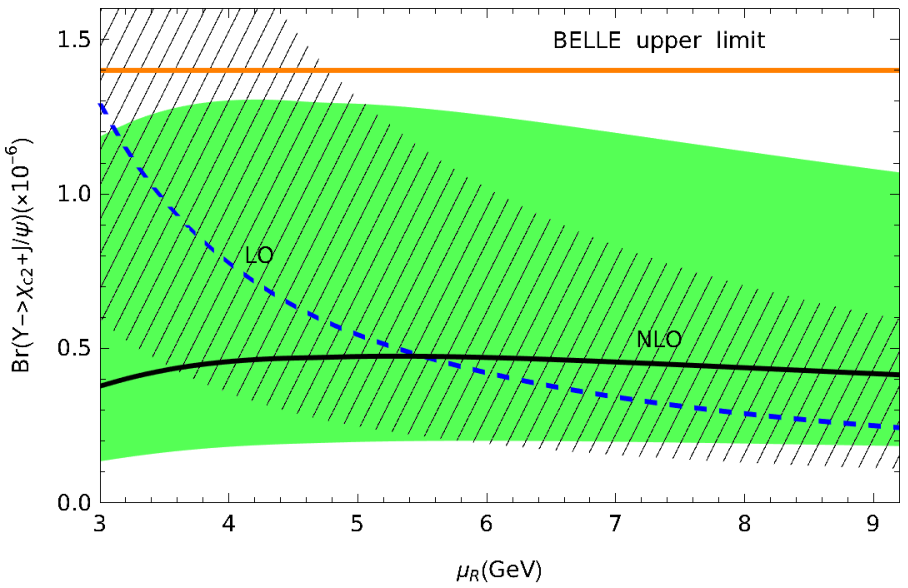}
\caption{
NRQCD results for ${\rm Br}(\Upsilon\to J/\psi+\eta_c(\chi_{cJ}))$ as functions of $\mu_R$.
The uncertainty bands for the theoretical results correspond to the choices of the charm and bottom quark mass. 
In addition, the uncertainty from the {\tt Belle} measurement for $J/\psi+\chi_{c1}$ is also illustrated.  
\label{fig-mu-dependence}}
\end{figure}

The results of the unpolarized decay widths at LO and NLO as functions of $\mu_R$ are presented in Fig.~\ref{fig-mu-dependence}.
For the sake of comparison, the {\tt Belle} measurements are also juxtaposed.
It is impressive that the renormalization scale dependence is significantly reduced for $J/\psi+\chi_{cJ}$ production, while slightly improved for $J/\psi+\eta_c$.
From Fig.~\ref{fig-mu-dependence}, we also observe that, after incorporating the $\mathcal{O}(\alpha_s)$ corrections,
the results for the decay width of the $J/\psi+\chi_{c1}$ production agrees perfectly with the {\tt Belle} measurement, albeit with large uncertainties.
For all the other processes, our NLO results are compatible with the upper limits given by experiment.

Finally, we list the results of the branching fractions for various processes in Tab.~\ref{tab-excited} and confront them with the {\tt Belle} data,
which may provide useful information for experimentalists. 
It can be found that most of the theoretical results are compatible with the measurements except for the process $\Upsilon(2S) \to J/\psi+\chi_{c1}$,
where some tension exists between theory and experiment.
Since all the predicted branching fractions are of order $10^{-6}$, given the high luminosity, it seems to be prospective to measure these processes at the super B-factory.

\begin{table}[h]\scriptsize
\caption{
Results of the branching fractions ($\times 10^{-6}$) for $\Upsilon\to J/\psi+\eta_c(\chi_{cJ})$.
The two uncertainties in the theoretical predictions are from the choices of the heavy quark mass and renormalization scale.
For comparison, the {\tt Belle} data~\cite{Belle:2014wam} is juxtaposed in the last row.
\label{tab-excited}}
\centering
\renewcommand{\arraystretch}{1.5}
\begin{tabular}{|c|c|c|c|}
\hline
Channels&LO&NLO&{\tt Belle}~\cite{Belle:2014wam}
\\
\hline
$\rm \Upsilon\to J/\psi+\eta_{c}$&$3.22^{+3.17+3.59}_{-1.58-1.96}$&$1.33^{+1.92+0.46}_{-0.78-0.95}$&$<2.2$
\\
\hline
$\rm \Upsilon\to J/\psi+\chi_{c 0}$&$1.42^{+1.36+2.34}_{-0.66-1.08}$&$1.36^{+1.81+0.003}_{-0.75-0.43}$&$<3.4$
\\
\hline
$\rm \Upsilon\to J/\psi+\chi_{c 1}$&$4.08^{+4.76+4.62}_{-2.08-2.55}$&$3.73^{+5.10+0.10}_{-2.06-1.19}$&$3.90^{+1.21+0.23}_{-1.21-0.23}$
\\
\hline
$\rm \Upsilon\to J/\psi+\chi_{c 2}$&$0.50^{+0.65+0.58}_{-0.27-0.31}$&$0.48^{+0.86+0.04}_{-0.29-0.12}$&$<1.4$
\\
\hline
$\rm \Upsilon\to J/\psi+\eta_{c}(2S)$&$2.10^{+2.07+2.34}_{-1.03-1.28}$&$0.87^{+1.25+0.30}_{-0.51-0.62}$&$<2.2$
\\
\hline
$\rm \Upsilon(2S)\to J/\psi+\eta_{c}$&$2.71^{+2.67+3.02}_{-1.33-1.65}$&$1.12^{+1.62+0.39}_{-0.66-0.80}$&$<5.4$
\\
\hline
$\rm \Upsilon(2S)\to J/\psi+\chi_{c 0}$&$1.19^{+1.15+1.97}_{-0.56-0.91}$&$1.15^{+1.52+0.003}_{-0.63-0.36}$&$<3.4$
\\
\hline
$\rm \Upsilon(2S)\to J/\psi+\chi_{c 1}$&$3.44^{+4.01+3.89}_{-1.75-2.15}$&$3.14^{+4.31+0.08}_{-1.74-1.00}$&$<1.2$
\\
\hline
$\rm \Upsilon(2S)\to J/\psi+\chi_{c 2}$&$0.42^{+0.55+0.49}_{-0.22-0.26}$&$0.41^{+0.73+0.04}_{-0.25-0.11}$&$<2.0$
\\
\hline
$\rm \Upsilon(2S)\to J/\psi+\eta_{c}(2S)$&$1.77^{+1.75+1.98}_{-0.87-1.08}$&$0.73^{+1.06+0.25}_{-0.43-0.53}$&$<2.5$
\\
\hline
$\rm \Upsilon(3S)\to J/\psi+\eta_{c}$&$3.27^{+3.22+3.64}_{-1.61-1.99}$&$1.35^{+1.95+0.46}_{-0.80-0.97}$&$-$
\\
\hline
$\rm \Upsilon(3S)\to J/\psi+\chi_{c 0}$&$1.44^{+1.38+2.38}_{-0.67-1.09}$&$1.38^{+1.84+0.003}_{-0.76-0.43}$&$-$
\\
\hline
$\rm \Upsilon(3S)\to J/\psi+\chi_{c 1}$&$4.14^{+4.83+4.69}_{-2.11-2.59}$&$3.79^{+5.19+0.10}_{-2.09-1.21}$&$-$
\\
\hline
$\rm \Upsilon(3S)\to J/\psi+\chi_{c 2}$&$0.51^{+0.66+0.59}_{-0.27-0.32}$&$0.49^{+0.88+0.05}_{-0.30-0.13}$&$-$
\\
\hline
\end{tabular}
\end{table}

In summary, the NLO perturbative corrections to $\Upsilon\to J/\psi+\eta_c(\chi_{cJ})$ is studied in the NRQCD framework.
It is the first time a process involving two-loop diagrams with six external legs is completely evaluated.
The decay widths for each specific helicity configuration are given,
and the polarization of the $J/\psi$ meson is also investigated.
At NLO, the renormalization scale dependence is significantly reduced for the $J/\psi+\chi_{cJ}$ production, while slightly improved for $J/\psi+\eta_c$. 
Our result of branching fraction for $\Upsilon\to J/\psi+\chi_{c1}$ is $3.73^{+5.10+0.10}_{-2.06-1.19}\times 10^{-6}$,
which agrees perfectly with the {\tt Belle} measurement, $(3.90\pm 1.21\pm0.23)\times 10^{-6}$. 
The theoretical predictions for other processes are compatible with the {\tt Belle} experimental limits,
except for $\Upsilon(2S)\to J/\psi+\chi_{c1}$ where there exists slight tension between theory and experiment.
Given the high luminosity at the super B-factory, it is prospective to measure the decay widths $\Upsilon$ to double charmonia as well as the $J/\psi$ polarization parameter,
which may provide crucial information for the QCD effective theories.

\begin{acknowledgments}
The work of Y.-D. Z. and W.-L. S. is supported by the National Natural Science Foundation
of China under Grants No. 11975187.  The work of H.-F. Zhang is supported by
the National Natural Science Foundation of China under Grants No. 11965006.
\end{acknowledgments}


\begin{thebibliography}{99}
\bibitem{Bodwin:1994jh}
G.~T.~Bodwin, E.~Braaten and G.~P.~Lepage,
Phys. Rev. D \textbf{51}, 1125-1171 (1995)
[erratum: Phys. Rev. D \textbf{55}, 5853 (1997)]
doi:10.1103/PhysRevD.55.5853
[arXiv:hep-ph/9407339 [hep-ph]].

\bibitem{Jia:2006rx}
Y.~Jia,
Phys. Rev. D \textbf{78}, 054003 (2008)
doi:10.1103/PhysRevD.78.054003
[arXiv:hep-ph/0611130 [hep-ph]].

\bibitem{Gong:2008ue}
B.~Gong, Y.~Jia and J.~X.~Wang,
Phys. Lett. B \textbf{670}, 350-355 (2009)
doi:10.1016/j.physletb.2008.10.063
[arXiv:0808.1034 [hep-ph]].

\bibitem{Braguta:2009xu}
V.~V.~Braguta and V.~G.~Kartvelishvili,
Phys. Rev. D \textbf{81}, 014012 (2010)
doi:10.1103/PhysRevD.81.014012
[arXiv:0907.2772 [hep-ph]].

\bibitem{Sun:2010qx}
P.~Sun, G.~Hao and C.~F.~Qiao,
Phys. Lett. B \textbf{702}, 49-54 (2011)
doi:10.1016/j.physletb.2011.06.060
[arXiv:1005.5535 [hep-ph]].

\bibitem{Braguta:2005gw}
V.~V.~Braguta, A.~K.~Likhoded and A.~V.~Luchinsky,
Phys. Rev. D \textbf{72}, 094018 (2005)
doi:10.1103/PhysRevD.72.094018
[arXiv:hep-ph/0506009 [hep-ph]].

\bibitem{Braguta:2009df}
V.~V.~Braguta, A.~K.~Likhoded and A.~V.~Luchinsky,
Phys. Rev. D \textbf{80}, 094008 (2009)
[erratum: Phys. Rev. D \textbf{85}, 119901 (2012)]
doi:10.1103/PhysRevD.80.094008
[arXiv:0902.0459 [hep-ph]].

\bibitem{Zhang:2011ng}
J.~Zhang, H.~Dong and F.~Feng,
Phys. Rev. D \textbf{84}, 094031 (2011)
doi:10.1103/PhysRevD.84.094031
[arXiv:1108.0890 [hep-ph]].

\bibitem{Sang:2011fw}
W.~L.~Sang, R.~Rashidin, U.~R.~Kim and J.~Lee,
Phys. Rev. D \textbf{84}, 074026 (2011)
doi:10.1103/PhysRevD.84.074026
[arXiv:1108.4104 [hep-ph]].

\bibitem{Chen:2012ih}
L.~B.~Chen and C.~F.~Qiao,
JHEP \textbf{11}, 168 (2012)
doi:10.1007/JHEP11(2012)168
[arXiv:1204.0215 [hep-ph]].

\bibitem{Jia:2007hy}
Y.~Jia,
Phys. Rev. D \textbf{76}, 074007 (2007)
doi:10.1103/PhysRevD.76.074007
[arXiv:0706.3685 [hep-ph]].

\bibitem{Sang:2015owa}
W.~L.~Sang, F.~Feng and Y.~Q.~Chen,
Phys. Rev. D \textbf{92}, no.1, 014025 (2015)
doi:10.1103/PhysRevD.92.014025
[arXiv:1502.01499 [hep-ph]].

\bibitem{Xu:2012uh}
J.~Xu, H.~R.~Dong, F.~Feng, Y.~J.~Gao and Y.~Jia,
Phys. Rev. D \textbf{87}, no.9, 094004 (2013)
doi:10.1103/PhysRevD.87.094004
[arXiv:1212.3591 [hep-ph]].

\bibitem{Belle:2014wam}
S.~D.~Yang \textit{et al.} [Belle],
Phys. Rev. D \textbf{90}, no.11, 112008 (2014)
doi:10.1103/PhysRevD.90.112008
[arXiv:1409.7644 [hep-ex]].

\bibitem{Zhang:2021ted}
Y.~D.~Zhang, F.~Feng, W.~L.~Sang and H.~F.~Zhang,
JHEP \textbf{12}, 189 (2021)
doi:10.1007/JHEP12(2021)189
[arXiv:2109.15223 [hep-ph]].

\bibitem{Feng:2019zmt}
F.~Feng, Y.~Jia and W.~L.~Sang,
[arXiv:1901.08447 [hep-ph]].

\bibitem{Sang:2022kub}
W.~L.~Sang, F.~Feng, Y.~Jia, Z.~Mo and J.~Y.~Zhang,
[arXiv:2202.11615 [hep-ph]].

\bibitem{Binosi:2008ig}
D.~Binosi, J.~Collins, C.~Kaufhold and L.~Theussl,
Comput. Phys. Commun. \textbf{180}, 1709-1715 (2009)
doi:10.1016/j.cpc.2009.02.020
[arXiv:0811.4113 [hep-ph]].

\bibitem{Haber:1994pe}
H.~E.~Haber,
[arXiv:hep-ph/9405376 [hep-ph]].

\bibitem{Jacob:1959at}
M.~Jacob and G.~C.~Wick,
Annals Phys. \textbf{7}, 404-428 (1959)
doi:10.1016/0003-4916(59)90051-X

\bibitem{Chernyak:1980dj}
V.~L.~Chernyak and A.~R.~Zhitnitsky,
Sov. J. Nucl. Phys. \textbf{31}, 544-552 (1980)

\bibitem{Brodsky:1981kj}
S.~J.~Brodsky and G.~P.~Lepage,
Phys. Rev. D \textbf{24}, 2848 (1981)
doi:10.1103/PhysRevD.24.2848

\bibitem{Hahn:2000kx}
T.~Hahn,
Comput. Phys. Commun. \textbf{140}, 418-431 (2001)
doi:10.1016/S0010-4655(01)00290-9
[arXiv:hep-ph/0012260 [hep-ph]].

\bibitem{Shtabovenko:2016sxi}
V.~Shtabovenko, R.~Mertig and F.~Orellana,
Comput. Phys. Commun. \textbf{207}, 432-444 (2016)
doi:10.1016/j.cpc.2016.06.008
[arXiv:1601.01167 [hep-ph]].

\bibitem{Feng:2012tk}
F.~Feng and R.~Mertig,
[arXiv:1212.3522 [hep-ph]].

\bibitem{Feng:2012iq}
F.~Feng,
Comput. Phys. Commun. \textbf{183}, 2158-2164 (2012)
doi:10.1016/j.cpc.2012.03.025
[arXiv:1204.2314 [hep-ph]].

\bibitem{Smirnov:2014hma}
A.~V.~Smirnov,
Comput. Phys. Commun. \textbf{189}, 182-191 (2015)
doi:10.1016/j.cpc.2014.11.024
[arXiv:1408.2372 [hep-ph]].

\bibitem{Patel:2015tea}
H.~H.~Patel,
Comput. Phys. Commun. \textbf{197}, 276-290 (2015)
doi:10.1016/j.cpc.2015.08.017
[arXiv:1503.01469 [hep-ph]].

\bibitem{Liu:2017jxz}
X.~Liu, Y.~Q.~Ma and C.~Y.~Wang,
Phys. Lett. B \textbf{779}, 353-357 (2018)
doi:10.1016/j.physletb.2018.02.026
[arXiv:1711.09572 [hep-ph]].

\bibitem{Liu:2020kpc}
X.~Liu, Y.~Q.~Ma, W.~Tao and P.~Zhang,
Chin. Phys. C \textbf{45}, no.1, 013115 (2021)
doi:10.1088/1674-1137/abc538
[arXiv:2009.07987 [hep-ph]].

\bibitem{Liu:2021wks}
X.~Liu and Y.~Q.~Ma,
Phys. Rev. D \textbf{105}, no.5, 5 (2022)
doi:10.1103/PhysRevD.105.L051503
[arXiv:2107.01864 [hep-ph]].

\bibitem{Liu:2022chg}
X.~Liu and Y.~Q.~Ma,
[arXiv:2201.11669 [hep-ph]].

\bibitem{Eichten:1995ch}
E.~J.~Eichten and C.~Quigg,
Phys. Rev. D \textbf{52}, 1726-1728 (1995)
doi:10.1103/PhysRevD.52.1726
[arXiv:hep-ph/9503356 [hep-ph]].

\bibitem{Chetyrkin:2000yt}
K.~G.~Chetyrkin, J.~H.~Kuhn and M.~Steinhauser,
Comput. Phys. Commun. \textbf{133}, 43-65 (2000)
doi:10.1016/S0010-4655(00)00155-7
[arXiv:hep-ph/0004189 [hep-ph]].

\bibitem{ParticleDataGroup:2020ssz}
P.~A.~Zyla \textit{et al.} [Particle Data Group],
PTEP \textbf{2020}, no.8, 083C01 (2020)
doi:10.1093/ptep/ptaa104

\end{thebibliography}
\end{document}